\documentstyle[12pt, aaspp4]{article}
\def \sun {$_{\scriptscriptstyle \odot}$}
\def \ltaprx {\lower .1ex\hbox{\rlap{\raise .6ex\hbox{\hskip .3ex
        {\ifmmode{\scriptscriptstyle <}\else
                {$\scriptscriptstyle <$}\fi}}}
        \kern -.4ex{\ifmmode{\scriptscriptstyle \sim}\else
                {$\scriptscriptstyle\sim$}\fi}}}
\def\gtaprx {\lower .1ex\hbox{\rlap{\raise .6ex\hbox{\hskip .3ex
        {\ifmmode{\scriptscriptstyle >}\else
                {$\scriptscriptstyle >$}\fi}}}
        \kern -.4ex{\ifmmode{\scriptscriptstyle \sim}\else
                {$\scriptscriptstyle\sim$}\fi}}}
\lefthead{FRYER \& WOOSLEY}
\righthead{}
\begin{document}
\begin{center} To be submitted to {\em The Astrophysical Journal}
\end{center}
\vspace{1.cm}
\title{Merging White Dwarf/Black Hole Binaries and Gamma-Ray Bursts} 
\author{Chris L. Fryer, S. E. Woosley}
\affil{Lick Observatory, University of California Observatories,
\\  Santa Cruz, CA 95064 \\ cfryer@ucolick.org}
\authoremail{cfryer@ucolick.org, woosley@ucolick.org}
\author{Marc Herant}
\affil{Washington University School of Medicine,
\\ Box 8107, 660 S. Euclid \\ St. Louis, MO 63110}
\authoremail{herantm@medicine.wustl.edu}
\author{Melvyn B. Davies}
\affil{Cambridge Institute for Astronomy,
\\ Madingley Road, Cambridge CH3 0HA\footnote{Now at:  Astronomy 
Group, University of Leicester, Leicester LE1 7RH}}
\authoremail{mbd@ast.cam.ac.uk}

\begin{abstract} 

The merger of compact binaries, especially black holes and neutron
stars, is frequently invoked to explain gamma-ray bursts (GRB's).  In
this paper, we present three dimensional hydrodynamical simulations 
of the relatively neglected mergers of
white dwarfs and black holes. During the merger, the white
dwarf is tidally disrupted and sheared into an accretion disk.  
Nuclear reactions are followed and the energy release is 
negligible.  Peak accretion rates are $\sim$0.05 M\sun\ s$^{-1}$ 
(less for lower mass white dwarfs) lasting for approximately a minute.  
Many of the disk parameters can be explained by a simple analytic model 
which we derive and compare to our simulations.  This model can be used 
to predict accretion rates for white dwarf and black hole (or neutron 
star) masses which are not simulated here.  Although the 
mergers studied here create disks with larger radii, and longer 
accretion times than those from the merger of double neutron stars, a 
larger fraction of the white dwarf's mass becomes part of the disk.  Thus
the merger of a white dwarf and a black hole could produce a 
long duration GRB.  The event rate of these mergers may be 
as high as $10^{-6}$\,yr$^{-1}$ per galaxy.

\end{abstract}


\section{Introduction}

As evidence supporting the extra-galactic nature of gamma-ray bursts
(GRB's) mounts (Metzger et al. 1997; Frail et al. 1997), the class of
models based on hyper-accreting black holes has become
the favorite mechanism for driving these explosions (e.g.,
Popham, Woosley, \& Fryer 1998; Eberl, Ruffert, \& Janka 1998).
Calculations show that a fraction of the gravitational potential
energy released as the material in the disk ($M_{\rm disk} \approx
0.01-2$M\sun) accretes into a small black hole ($M_{\rm BH} \approx
3-10$M\sun) can be converted into a ``fireball'' which produces the
observed gamma-rays (Meszaros \& Rees 1992).  Such systems form in
collapsars or hypernovae (Woosley 1993, 1996; Paczynski 1997) and in the
merger of compact binaries consisting of: two neutron stars or a
neutron star and a black hole (Paczy\'nski 1991; Narayan, Paczy\'nski, 
\& Piran 1992); a helium star and a black hole (Fryer \& Woosley 1998), 
and, the topic of this paper, a white dwarf and a black hole.

Several mechanisms have been proposed to facilitate the conversion of
potential energy into GRB explosion energy.  Due to the high densities
involved during the merging process, the potential energy may be
emitted in the form of neutrinos, and the subsequent annihilation
of these neutrinos could power a GRB (Goodman, Dar, \& Nussinov 1987).
Meszaros \& Rees (1992) first pointed out the advantages of disk
geometry for enhancing neutrino annihilation. However, unless the
accretion rate is very high, over a few hundredths of a solar mass per
second, Popham et al. (1998) have shown that neutrino emission is 
inefficent as the energy released in the disk is advected into the hole.
Alternatively, and especially for lower accretion rates and lower
disk viscosity, the amplification of magnetic field in the disk 
can tap either the black hole rotational energy or the potential 
energy of the accreting material to drive
to drive a relativistic jet and create a GRB (Blandford \& Znajek 
1977; MacDonald et al. 1986; Paczynski 1991,1997; Woosley 1996; 
Meszaros \& Rees 1997, Katz 1997).  However, these magnetically 
based models are currently not sufficiently accurate to 
make quantitative predictions (Livio, Ogilvie, \& Pringle 1998).   

One scenario to form binaries consisting of a black hole and a 
white dwarf (WD/BH binaries) begins with main-sequence stellar systems 
having extreme mass ratios ($M_{\rm primary} \gtaprx 30$M\sun, 
$M_{\rm secondary} \approx 1-8$M\sun).  The formation scenario for white 
dwarf binaries with neutron star companions is similar, only with a 
primary star mass  between $\approx$8 and $\approx$30M\sun.  
As the massive star evolves off the main 
sequence, a common envelope phase may occur that ejects the 
primary's hydrogen envelope.  Beyond some critical mass (roughly 
$30$M\sun), massive stars are thought to form black holes, 
either by a failed explosion, or through significant fallback 
(Woosley \& Weaver 1995, Fryer 1998).  The more massive primary 
eventually collapses into a $3-15$M\sun black hole and forms a 
binary consisting of a black hole and a main-sequence star.  
As the secondary expands and the orbit shrinks, a second mass transfer 
phase commences.  This phase is observed in the closest of these 
systems as a low-mass X-ray binary (e.g. J0422+32-Nova Per, 
2023+338 Nova Cyg).  Roughly 20 binary systems with a black hole and a 
low-mass companion have been observed
and many, as yet undetected systems may exist (see Tanaka \& Shibazaki,
1996, for a review).  These systems generally involve main-sequence
secondaries that lose a significant amount of their mass while still
burning hydrogen and do not evolve into massive ($> 0.5$M\sun)
white dwarfs.  However, slightly wider binaries do not undergo mass
transfer until after the secondary has evolved off the main sequence
and produce the systems which we model in this paper.
Unfortunately, the formation rate of the observed low mass X-ray
binaries is difficult to determine, and estimating the number 
of wider systems from observations is impossible.  It is these more 
massive companions that dominate the merging WD/BH binaries 
and hence, the typical merging system consists of a black hole 
and massive ($\gtrsim 0.9$M\sun) white dwarf.

An alternative evolution scenario for WD/BH binaries begins with less
extreme mass ratios in a system where the primary initially forms a neutron 
star in a supernova explosion.  As the secondary expands off the 
main-sequence and a common envelope phase ensues, the neutron star 
accretes rapidly via neutrino emission (allowing the accretion rate 
to greatly exceed the photon Eddington limit) and eventually collapses to a 
low mass black hole (see Fryer, Benz, \& Herant 1996, Bethe \& Brown 1998).  
Because this common envelope phase happens quickly, it is unlikely that 
systems evolving through this scenario will be observed, and no merger 
rate can be predicted from the observations.  

Unfortunately, uncertainties in black hole formation and binary 
evolution also make it difficult to make any firm predictions 
using population synthesis studies, but the merger rate is likely 
to lie in the range from $10^{-9}-10^{-6}$\,yr$^{-1}$ 
per Milky-Way like galaxy (Fryer, Woosley, \& Hartmann 1999).  The 
large uncertainties in the merger rate are primarily due 
to uncertainties in the critical mass beyond which massive 
progenitor stars collapse to black holes and in the kick 
imparted to black holes.  

Black holes and white dwarfs can also merge through collisions in
dense star regions such as in galactic centers and globular clusters.
Sigurdsson \& Rees (1997) predict a neutron star/white dwarf merger
rate of $\sim 10^{-7}$\,yr$^{-1}$ per galaxy.  Low mass black holes
will merge with white dwarfs roughly at the same rate (or an order of
magnitude less) depending upon the black hole formation rate.  This
rate is comparable to the merger rate predicted by Quinlan \& Shapiro
(1987, 1989, 1990).  Depending on the beaming fraction, this rate is
certainly sufficient to give the observed GRB statistics (Wijers et
al. 1998), provided, of course, that the merger produces a GRB.

In this paper, we model the merger of black holes and white dwarfs on
the computer using a three-dimensional hydrodynamics code based on the
Smooth Particle Hydrodynamics (SPH) method (Benz 1990). 
We follow the merger from the initial
Roche-lobe overflow through the complete disruption of the white dwarf 
into a disk.  Roche-lobe overflow for compact objects differs from that 
for giant stars in several ways:  (b) due to the degeneracy 
of the compact object, its radius increases as it loses mass, and 
(c) the orbital angular momentum is far from conserved ($\equiv$ the 
mass transfer is not ``conservative'').  We discuss this physics, applicable 
to most compact object mergers, in \S 2.  A description of the code 
along with a presentation of the simulations, including a comparison to the
analytic estimates of \S 2, is given in \S 3.  We conclude with a
discussion of the accretion disks formed in these mergers and their 
suitability as GRB models.

\section{Accretion Disk Formation}

Whether or not the accretion disks formed in WD/BH mergers 
produce the necessary GRB explosion energies is determined by the
accretion rate and the angular momentum of those disks, which in turn,
depends upon the size and mass of the accretion disk.  It is important
to know, then, how quickly, and at what radius, the white dwarf is
torn up by the gravitational potential of the black hole and
transformed into an accretion disk which might fuel a GRB.  One might
naively assume that, since the white dwarf is less massive than the
black hole ($M_{\rm BH} \sim 3-10$M\sun, $M_{\rm WD} \sim 0.5-1.3$M\sun), 
that stable accretion will occur and the white dwarf will
slowly accrete onto the black hole over many orbital periods.
However, as we shall discuss in this section, several aspects of physics
conspire to destabilize this mass transfer, leading to the rapid
transformation of most of the white dwarf into an accretion disk.  
This was seen in the merger of double white dwarf binaries 
by Davies, Benz \& Hills (1991). 

\subsection{Gravitational Radiation}
Gravitational radiation plays an important role in the merging 
of double neutron star or black hole/neutron star systems.  For these
systems, the emission of gravitational waves tightens the binaries 
on a timescale comparable to those of the hydrodynamical evolution.  
However, white dwarfs fill their Roche lobes at much wider separations where
the gravitational wave merging timescale is 1-100\,yr 
(depending upon the white dwarf and black hole masses).  
Although gravitational radiation does cause the orbit to tighten 
sufficiently to drive the white dwarf to fill its
Roche lobe in the first place, once Roche-lobe overflow occurs, the 
mass transfer rate from the white dwarf onto a disk around the black 
hole is determined by the transfer of angular momentum and the white dwarf 
mass-radius relationship which drive unstable mass transfer on much 
shorter timescales ($\sim$minutes).

\subsection{Effects of Degeneracy}
One such destabilizing effect is the inverse relationship of the radii of 
degenerate objects (neutron star, white dwarf) with respect to mass.
For a $\Gamma=5/3$ polytrope approximation 
of a white dwarf equation of state, this relationship is (Nauenberg 1972):
\begin{equation}\label{eq:rwd}
R_{\rm WD} \approx 
10^4 \left(\frac{M_{\rm WD}}{0.7 {\rm M_\odot}} \right)^{-1/3}
\left[1-\left(\frac{M_{\rm WD}}{M_{\rm CH}} \right)^{4/3} \right]^{1/2} \,
\left( \frac{\mu_e}{2} \right)^{-5/3} \,{\rm km}
\end{equation}
where $R_{\rm WD}$, $M_{\rm WD}$, and $\mu_e$ are the radius, 
mass, and mean molecular weight per electron of the white dwarf 
and $M_{\rm CH}\approx 1.4$ is the Chandrasekhar mass.  
Our simulated white dwarfs differ slightly from this simple 
relation due to deviations from a simple $\Gamma=5/3$ 
polytrope (Fig. 1).  The radius of the white dwarf and the 
masses of the white dwarf and black hole determine the orbital 
separation at which Roche-lobe overflow commences (Eggleton 1983):
\begin{equation}\label{eq:rl}
A_0=R_{\rm WD} \frac{0.6 q^{2/3}+{\rm ln} (1+q^{1/3})}{0.49 q^{2/3}}
\end{equation}
where $q=M_{\rm WD}/M_{\rm BH}$ is the mass ratio (Fig. 1).  
Stable mass transfer would require that as the white dwarf loses 
mass, its orbit widens to place it just at this critical 
Roche lobe separation.  If the white dwarf binary instead remains 
at a constant orbital separation, the accretion will quickly become 
unstable as the white dwarf itself expands.  This effect is important 
for all merging systems involving a compact secondary.

\subsection{Non-Conservative Mass Transfer}
The orbital separation need not remain constant.  In conservative
mass transfer, when an object accretes onto a more massive companion,
orbital angular momentum conservation requires that the orbit expands.
However, some fraction of the material can be lost from the system and
carry away angular momentum.  Hence, although the total angular
momentum is conserved, the orbital angular momentum of the binary
system decreases, and the orbital separation may actually decrease
during mass transfer.  This ``non-conservative'' mass-transfer can be
parameterized and solved (see Podsiadlowski, Joss, \& Hsu 1992 and
references therein).  In addition, for mergers with black holes or 
neutron stars, some of the orbital angular momentum is converted to 
angular momentum of the accretion disk or to spin angular momentum 
of the black hole.  The change of orbital angular momentum 
($\delta J_{\rm orbit}$) of the binary is then given by: 
\begin{equation} 
\delta J_{\rm orbit} = \left[j_{\rm ejecta} (1-\beta)+ j_{\rm disk} 
\beta \right] \delta M_{\rm WD} \frac{2 \pi A^2}{P} 
\end{equation} 
where $\beta$ is the fraction of mass lost by the white dwarf that is
accreted by the black hole (or becomes part of the black hole's
accretion disk), $j_{\rm ejecta}$ and $j_{\rm disk}$ are the specific
angular momenta (in the rest frame of the black hole) 
of the ejected material and the material which is either
accreted onto the black hole or becomes part of the accretion disk.
$A$ and $P$ are the orbital separation and period of the binary
system.  Following the procedure of Podsiadlowski, Joss, \& Hsu
(1992), we derive the orbital separation ($A$) of the binary during
mass transfer including the loss of angular momentum to the accretion
disk: 
\begin{equation}
\label{eq:rad} 
\frac{A}{A^0}=\frac{M_{\rm WD}+M_{\rm BH+disk}}{M_{\rm WD}^0+M_{\rm
BH}^0} \left( \frac{M_{\rm WD}}{M_{\rm WD}^0} \right)^{C_1} 
\left ( \frac{M_{\rm BH}}{M_{\rm BH}^0} \right)^{C_2} 
\end{equation} 
where the values of the constants differ only slightly from those
derived by Podsiadlowski, Joss, \& Hsu (1992): 
\begin{equation} 
C_1 \equiv 2 j_{\rm ejecta} (1-\beta) - 2 + 2 j_{\rm disk} \beta
\end{equation} 
\begin{equation} 
C_2 \equiv \frac{-2 j_{\rm ejecta}}{\beta} (1-\beta) - 2 - 2 j_{\rm disk} 
\end{equation} 
and
\begin{equation} 
M_{\rm BH+disk}=\beta (M_{\rm WD}^0-M_{\rm WD}) + M_{\rm BH}^0, 
\end{equation} 
and where superscript 0 denotes pre-mass transfer phase values.

In Roche-lobe overflow onto compact objects (neutron stars or black holes), 
much of the angular momentum is placed into a disk around that compact 
object.  For the merger of binaries consisting of a black hole and 
a neutron star, roughly half of the orbital angular momentum is fed 
directly into spinning up the black hole (Eberl, Ruffert, \& Janka 1998).  
For wider Roche-lobe overflow systems (e.g. white dwarf mergers) much of 
the orbital angular momentum is converted into disk angular momentum.
(see Papaloizou \& Lin 1995 for a review).  In systems where the 
mass transfer is stable and the primary, disk and secondary coexist for 
many orbital periods, the angular momentum of the disk can be transferred 
back to the orbital angular momentum of the binary.  However, for the 
runaway accretion caused by the expansion of the white dwarf, the white 
dwarf is disrupted quickly (2-3 orbital periods) and the torques between 
the disk and the binary stars are unable convert the disk angular 
momentum back to that of the orbit before the disruption of the 
white dwarf.  

Figure 2 shows the orbital evolution for $0.7$ and 
$1.1$M\sun white dwarfs merging with a $3$M\sun
black hole for a range of values of $j_{\rm disk}$ (in terms of the 
white dwarf specific angular momentum $\equiv j_{\rm WD}$) and assuming no
mass is ejected from the system ($\beta=1$).  The critical separation
for Roche lobe overflow, shown in Figure 2, marks the dividing line
between stable and unstable mass accretion.  If the orbit widens
faster than the white dwarf expands, the accretion rate onto the black
hole is limited to the gravitational wave timescale (1-100\,yr) and
the merger occurs on these timescales.  However, if $j_{\rm
disk}>0.3,0.1 j_{\rm WD}$ for $0.7,1.1$M\sun white dwarfs respectively, the
mass transfer is unstable.  If the accreting material transports all
of its angular momentum to the accretion disk, then the angular 
momentum of the disk is roughly the angular momentum of the 
material at the Lagrange point.  Assuming tidal locking, for 
the binary systems we model, this angular momentum is roughly: 
$j_{\rm disk} \approx (A-R_{\rm WD})^2/A^2 j_{\rm WD} \sim 0.5 
j_{\rm WD}$ where $A$ is the orbital separation, $R_{\rm WD}$ 
is the white dwarf radius, and $j_{\rm WD}$ is the specific angular momentum 
of the white dwarf in the rest frame of the black hole ($\sim 10^{18} 
{\rm cm^2 \, s^{-1}}$).  For these high values of 
$j_{\rm disk}$, unstable mass transfer is inevitable, and we expect 
the white dwarf to be tidally disrupted rapidly.  But to accurately 
calculate the mass transfer, and ultimately, the mass accretion 
rate onto the black hole, we must resort to numerical simulations.

\section{Simulations}

For our simulations, we use a three dimensional SPH code
(Davies, Benz, \& Hills 1991) with 6000-16000 particles.  We employ
the equation of state developed by Lattimer \& Swesty (1991) for
densities above $10^{11}\,{\rm g \, cm^{-3}}$ and, for low densities,
the equation of state by Blinnikov, Dunina-Barkovskaya \& Nadyozhin
(1996).  We include a nuclear burning network for temperatures
above $4\times10^8\,{\rm K}$ (Woosley 1986), though we find that 
burning is not important, except well within the accretion disk.
As we are concerned with the tidal disruption of the white dwarf and 
not the accretion of matter in the disk formed from this disruption,
we model the black hole (or neutron star) as a point mass and remove 
particles that fall within $2-3 \times 10^8$\,cm of the black hole, 
well before general relativistic effects are important.  Similarly, 
since we are not following the evolution of the accretion disk, the 
the numerically determined artificial viscosity should not impact 
our results.  However, as a check, we have varied the artificial 
viscosity by an order of magnitude and find it does not effect the 
radius at which the white dwarf is disrupted or the initial structure 
of the accretion disk formed by this disruption.  On the other hand, 
the true physical viscosity does affect the rate at which material is 
accreted onto the black hole, and hence the pair fireball energy, 
which we discuss in \S 4.

With this code, we modeled the tidal disruption of 4 binary 
systems consisting of a white dwarf (with masses of 
0.7, 1.1\,M\sun) and a black hole (with masses 
of 3, 10\,M\sun) and one system consisting of a $1.1$\,M\sun 
white dwarf and a $1.4$\,M\sun neutron star.  We followed the 
evolution from the initial Roche-lobe overflow through the 
destruction of the white dwarf and the formation of an 
accretion disk (Figs. 3, 4).  We assume that gravitational 
radiation has brought the white dwarf close enough to its 
black hole companion to overfill its Roche-lobe and transfer 
mass onto the black hole.  We estimate this critical 
separation using eq. (\ref{eq:rl}).  By increasing the 
separation by 20-30\%, we see that no mass transfer 
takes place (Fig. 3), assuring that our initial separation 
is within 30\% of the actual Roche-lobe overflow separation.  
We will come back to this error estimate in our discussion of the 
accretion disk properties at the end of this section.

Before we discuss the disk properties, let us first validate our physical 
picture of the tidal disruption process.  From \S 2, we expect the specific 
angular momentum of the disk (in the rest frame of the black hole) to be 
initially $\sim 0.5 j_{\rm WD}$ and then rise as more of the white dwarf 
it disrupted.  The angular momentum of the simulated disk is 
$\sim 0.6 j_{\rm WD}$ and then increases to $1.0 
j_{\rm WD}$ as the white dwarf is disrupted (Fig. 5).  Physically, this means 
that as the white dwarf transfers mass onto a disk around the 
black hole, the angular momentum of matter at the Lagrange point is first 
added to the disk.  When the white dwarf is finally disrupted, 
nearly all of its angular momentum is immediately put into the 
disk, and the average disk angular momentum equals the initial 
white dwarf angular momentum.  The disk must then shed this angular 
momentum before this material can accrete onto the black hole (see \S 4).

Because much of the orbital angular momentum is converted 
into disk angular momentum, the orbital separation does not 
expand as one might expect in conservative mass-transfer, 
and the white dwarf is quickly disrupted by tidal forces.
As the black hole accretes mass, the orbital separation 
of the white dwarf/black hole binary is described by 
equation (\ref{eq:rad}).  Using equation (\ref{eq:rad}) 
and assuming no mass is ejected from the system 
(very little mass is ejected in our simulations, see Figs. 3, 4), 
we can plot data from the simulations along with the derived 
separations for a range $j_{\rm disk}$ values (Fig. 6).  The 
remarkable agreement of the best fit of $j_{\rm disk}$ 
using equation (\ref{eq:rad}) and the actual $j_{\rm disk}$ 
values from Figure 5 suggests that we have indeed 
found the relevant physics, and that the orbital separation 
can be estimated by our simple mass-transfer model.

In these simulations, the mass transfer from the white dwarf 
becomes increasingly unstable as more of the white dwarf 
expands beyond its Roche radius and accretes onto a disk 
around the black hole.  Our simulations show that after 
losing $\sim 0.2$M\sun, the transfer rate becomes so 
great that the white dwarf is disrupted.  This 
occurs rapidly (in an orbit time), dumping 
the remains of the  white dwarf into an accretion disk around 
the black hole.  This critical mass loss after which the accretion 
runs away is the one parameter not determined by our analytic model.  
Using our simulations to constrain this parameter, we are 
able to describe both the angular momentum 
and the mass growth rate of the disk from the tidal 
disruption of the white dwarf.

The specific angular momentum of the disk is given by
\begin{equation}
j \approx \sqrt{G A (M_{\rm BH}+M_{\rm WD}^{\rm disruption})}
\end{equation}
where $G$ is the gravitational constant, $M_{\rm BH}$ is the 
black hole mass, and $M_{\rm WD}^{\rm disruption} \approx
0.5,0.8 \,$M\sun (for initial white dwarf masses of 0.7, 1.1\,M\sun 
respectively) is the white dwarf mass 
at the time of disruption taken from our simulations.  
The orbital separation ($A$) can be derived from 
equation (\ref{eq:rad}).  The mass transfer rate of the white 
dwarf onto the disk is roughly 
\begin{equation}
\dot{M}=M_{\rm WD}^{\rm disruption}/T_{\rm orbit}
\end{equation}
where $T_{\rm orbit}$ is the orbital timescale for the binary system 
after the white dwarf has lost $0.2$M\sun.  These results 
are summarized in Table 1 and the mass-transfer rate can be 
compared to the simulated rates shown in Figure 7.  Note 
that mass-transfer rates derived from analytical estimates agree 
within a factor of 2 with those obtained from our simulations.
The actual accretion rate onto the black hole is not likely to 
exceed this mass-transfer rate.

Many of these results rely upon our knowing the exact separation 
where Roche lobe overflow commences.  As we have 
already mentioned, by increasing the separation by 30\%, we find 
no accretion occurs over many orbits, which suggests that the 
error in the initial separation is less than 30\%.  
If the errors in the initial orbital separation are less than 30\%, 
our maximum mass-transfer rates are accurate to $\lesssim$30\% and the 
disk angular momenta are accurate to $\lesssim$15\%.  Even changing 
the initial separation by a factor of 2 only results in a factor of 3 
change in the maximum accretion rate and a change in the angular 
momenta by less than 40\%.

\section{Accretion Disk Powered Gamma-Ray Bursts}

With these results, we can now address the viability of WD/BH
mergers as a GRB model.  The mass transfer rate of the white dwarf
onto the black hole accretion disk should, in a steady state, balance the
accretion rate into the black hole.  The actual accretion rate is 
determined by the efficiency at which the angular momentum is removed 
from the disk\footnote{This equation applies only when the scale 
height of the disk is roughly equal to the disk radius.  We may be 
underestimating the viscous timescale by an order of magnitude.  
However, we are most interested in deriving a lower limit for this 
timescale (upper limit for the accretion rate).} 
(Popham, Woosley,\& Fryer 1998):
\begin{equation}\label{eq:mdot}
\dot{M}_{\rm acc} \approx 0.37 \alpha M_{\rm disk} M_{\rm BH}^{1/2} 
r_{\rm disk,9}^{-3/2} \, \, {\rm M_\odot} \, \, {\rm s^{-1}},
\end{equation}
where $\alpha$ is the standard accretion disk parameter, 
$M_{\rm disk}$ and $M_{\rm BH}$ are, respectively, the mass of the disk 
and the black hole in M\sun, and $r_{\rm disk,9}$ is the outer 
disk radius in $10^9$\,cm.  Figure 8 shows the mass of the disk as a 
function of radius for
our two $M_{\rm BH}=3$M\sun models, from which, given a value of
$\alpha$, we can determine the accretion rate onto the black
hole.  For values of $\alpha<0.5$, the accretion rate is limited 
by the disk accretion and not the mass-transfer rate.  Using 
equation (\ref{eq:mdot}), we estimate the effective 
disk viscosity ($\alpha$) from accretion rate onto the black hole 
of our hydrodynamical simulations to be $\sim 0.1$. 

The energy from neutrino annihilation can be estimated by integrating 
the following approximate fit to the pair luminosity results of 
Popham, Woosley, \& Fryer (1998):
\begin{equation}
{\rm log}\,L_{\nu,\bar{\nu}}({\rm erg \, s^{-1}}) 
\approx 43.6+4.89\,{\rm log} \left(\frac{\dot{M}}
{0.01 {\rm M_\odot} {\rm \, s^{-1}}}\right)+3.4a
\end{equation} 
where $a \equiv J_{\rm BH} c/GM_{\rm BH}^2$ is the spin parameter.  
This fit is reasonably accurate for accretion rates between $0.01$ 
and $0.1$M\sun\,s$^{-1}$.  
Table 2 gives the maximum energies for each of our simulations.  
In the optimistic situation where $\alpha>0.5$ and the disk accretion 
rate equals the mass-transfer rate from the white dwarf into a 
black hole accretion disk, the disruption of a 
white dwarf around a black hole cannot explain the most energetic gamma-ray 
bursts without requiring that the mechanism produce strongly beamed 
jets.  Indeed, with isotropic energy requirements as high as
$3\times10^{53}\,\rm{erg}$ (Kulkarni et al. 1998), the beaming must be
extremely high (the burst must be constrained to 0.1\% of the sky, that 
is, the beaming factor $>1000$).  However, a wide range of GRB 
energies may exist, and WD/BH mergers may only constitute a subset of the
observations.  If $\alpha=0.1$, the accretion rate drops 
by about a factor of 5.  For the most optimistic mergers of a 
$1.1$M\sun white dwarf with a black hole, this lowers the accretion 
rate on the black hole to $0.01-0.02 {\rm M_\odot \, s^{-1}}$ 
and increases the accretion time, causing a net decrease in 
the total energy produced by neutrino annihilation of 
roughly 1-2 orders of magnitude.  With beaming factors of 
$\sim 100$, WD/BH mergers could still explain bursts with 
inferred isotropic energies between $10^{48}-10^{51}$\,erg.  

Alternatively, and perhaps more likely for the low-mass accretion 
rates derived here, the GRB can be powered by the magnetic fields 
of the disk, which become stretched and amplified as the material 
accretes.  These magnetic fields then extract the rotational energy 
of the black hole (Blandford \& Znajek 1977; MacDonald et al. 1986; 
Paczynski 1991,1997; Woosley 1993; Katz 1994, 1997; 
Hartmann \& Woosley 1995; Thompson 1996;  Meszaros \& Rees 1997; 
Popham et al. 1998).  Very roughly, using Blandford-Znajek for example,
\begin{equation}
L_{\rm rot}=10^{50} \left ( \frac{j c}{G M_{\rm BH}}\right )^2
\left ( \frac{M_{\rm BH}}{3 {\rm M_\odot}}\right )^2
\left ( \frac{B}{10^{15} {\rm Gauss}}\right )^2 \, {\rm erg\,s^{-1}}
\end{equation}
where $j$ is the specific angular momentum of the black hole and $B$
is the magnetic field strength in the disk.  Table 2 lists the total 
energy that an initially non-rotating black hole would emit over its 
accretion timescale assuming the magnetic field energy is 10\% of the 
equipartition energy, or $0.1 \rho v^2$.
These high magnetic fields are reasonable if the disk viscosity 
depends upon the magnetic field strength.  In this case,
the viscosity is initially $\sim 0$ allowing the disk to continue 
winding the magnetic field until a sufficiently strong equipartion 
field is generated, thereby increasing the viscosity and allowing the 
disk to accrete.  

A successful GRB explosion must also avoid excessive baryonic 
contamination.  The disruption of the white dwarf forms a 
hot thick disk around the black hole (Figure 9), with some of the 
matter along the angular momentum axis above the black hole (Table 2).
The explosion will force its way along this polar region, 
sweeping up this material (and possibly pushing some aside).  
Assuming all of the material 
is swept along with the burst, we can estimate a lower limit for 
the Lorentz factors (Table 2).  Beaming factors of at least 100 
are required to achieve the high Lorentz factors needed to power 
a gamma-ray burst.  Even assuming that beaming factors of 100 
do occur, low mass white dwarfs do not produce enough energy 
(or high enough Lorentz factors) to power a gamma-ray burst.  
Thus, there is some critical white dwarf mass (between 
$0.7-1.1$M\sun depending upon beaming) 
below which no visible GRB will form.  Because of the strong 
dependence of the GRB luminosity on the Lorentz factor, 
the transition from observed gamma-ray burst to non-detectable 
explosion is sharp.  Those explosions that do not achieve the 
high Lorentz factors will only be observable in our own Galaxy, 
and, given the low event rate, will not be detected.  
We reiterate, however, that most of the merging 
white dwarfs will be massive (Fryer, Woosley, \& Hartmann 1999) 
and a large fraction of the merging systems may become GRBs.

The merger of a black hole and a massive white dwarf can produce 
the energies ($10^{48}-10^{51} {\rm erg}$) and the high Lorentz factors 
to explain the long duration GRBs if the bursts themselves are highly 
beamed (beaming factors $>100$).  Assuming the GRB rate for isotropic 
bursts is $10^{-7}$\,yr$^{-1}$ per galaxy of roughly the Milky Way's 
size (Wijers et al. 1998), the merger rate of massive white dwarfs and 
black holes with beaming factors $>10$ must be $\gtrsim 
10^{-6}$\,yr$^{-1}$ per galaxy, within the uncertainties of the 
predicted rates (see \S 1).  

From these results, one might conclude that mergers of white dwarfs and 
black holes are not likely to play a major role in the production of 
gamma-ray bursts.  However, our estimates of the energy released via 
magnetic fields are very uncertain.  Some magnetic field mechanisms may 
convert a large fraction of the potential energy of the accreting material 
into burst energy.  If a magnetic field mechanism can be constructed 
which converts 10\% of the potential energy into burst energy, WD/BH 
mergers would have energies in excess of $10^{52}$erg.  With beaming 
into 10
gamma-ray bursts.  

The merger of a neutron star and a white dwarf is a different story.  
At these accretion rates, Popham, Woosley, \& Fryer (1998) found that 
much of the energy is advected into the black hole.  The hard surface 
of the neutron star acts as a plug, stopping up this accretion.  
Unless the neutron star mass quickly exceeds the upper neutron 
star mass limit, causing it to collapse to a black hole and removing 
this plug, the accreting material will flow around the neutron star, 
building up a spherically symmetric atmosphere.  Any 
explosion from the surface will be baryon rich with velocities 
much less than the speed of light (Fryer, Benz, \& Herant 1994).  
These outbursts will be too dim to observe beyond our Galaxy, 
and are too rare to observe within our Galaxy.  

\acknowledgements
This research has been supported by NASA (NAG5-2843 and MIT SC A292701), 
and the NSF (AST-97-31569).  We would like to thank Bob Popham, 
Thomas Janka, and William Lee for many useful corrections and comments.  
We acknowledge many helpful conversations and communications 
on the subject of gamma-ray bursts with Andrew MacFadyen, 
Dieter Hartmann, Max Ruffert, Jonathon Katz and we thank Aimee 
Hungerford for helpful comments on the manuscript.

{}
\newpage

\begin{deluxetable}{lccccc}
\tablewidth{38pc}
\tablecaption{WD/BH Mergers}
\tablehead{ \colhead{Disk Parameters} & 
\colhead{$M_{\rm WD}=0.7$} &
\colhead{$M_{\rm WD}=0.7$} &
\colhead{$M_{\rm WD}=1.1$} & \colhead{$M_{\rm WD}=1.1$} &
\colhead{$M_{\rm WD}=1.1$} \\
& \colhead{$M_{\rm BH}=10$} & \colhead{$M_{\rm BH}=3$} &
\colhead{$M_{\rm BH}=10$} & \colhead{$M_{\rm BH}=3$} &
\colhead{$M_{\rm NS}=1.4$}}

\startdata
$A(10^9\,{\rm cm})$ & 4.98 & 3.10 & 2.37 & 1.49 & 1.27\nl
$T_{\rm orbit}$(s) & 58.4 & 48.1 & 18.9 & 15.4 & 12.1 \nl
$j_{\rm disk}(10^{18}\,{\rm cm^2 \, s^{-1}})$ & 
2.67 & 1.24 & 1.87 & 0.903 & 0.651 \nl
$\dot{M}({\rm M_\odot} \, {\rm s^{-1}})$ 
& 0.00856 & 0.01 & 0.0477 & 0.0584 & 0.074 \nl
$A^{\rm sim}(10^9\,{\rm cm})\tablenotemark{a}$ & 
5.0 & 3.2 & 2.5 & 1.6 & 1.2 \nl
$M^{\rm disr}_{\rm WD}({\rm M_\odot})$\tablenotemark{b} 
& 0.5 & 0.5 & 0.8 & 0.8 & 0.8 \nl
$\dot{M}_{\rm peak}^{\rm sim}({\rm M_\odot} \, 
{\rm s^{-1}})$\tablenotemark{c} & 
0.012 & 0.008 & 0.063 & 0.079 & 0.075 \nl

\tablenotetext{a}{The last three rows of data come 
directly from the simulations, the other rows are 
derived from the equations in \S 2,3.}
\tablenotetext{b}{This is the simulated white dwarf 
mass at the onset of the white dwarf disruption.  For 
our derivations, we assume $M_{\rm WD}^{\rm disr}\approx
0.5,0.8$M\sun for initial white dwarf masses of 0.7,1.1 M\sun 
respectively.}
\tablenotetext{c}{The peak mass-transfer rate from 
Figure 5.}

\enddata
\end{deluxetable}

\begin{deluxetable}{lcccc}
\tablewidth{35pc}
\tablecaption{Powering a GRB}
\tablehead{ \colhead{Observables} & 
\colhead{$M_{\rm WD}=0.7$} &
\colhead{$M_{\rm WD}=0.7$} &
\colhead{$M_{\rm WD}=1.1$} &
\colhead{$M_{\rm WD}=1.1$} \\
& \colhead{$M_{\rm BH}=10$} & \colhead{$M_{\rm BH}=3$} &
\colhead{$M_{\rm BH}=10$} & \colhead{$M_{\rm BH}=3$}}

\startdata
$a \equiv jc/GM_{\rm BH}$ & 0.21 & 0.54 & 0.31 & 0.69 \nl
$E_{\nu,\bar{\nu}}^{\rm max} (10^{49}{\rm Erg})$\tablenotemark{a} 
& 0.001 & 0.003 & 3 & 50 \nl
$E_{\rm rot}(10^{49}{\rm erg})$\tablenotemark{b} 
& $\sim 0.1$ & $\sim 1$ & $\sim 1$  & $\sim 4$ \nl 
$M_{\rm axis} ({\rm M_\odot})$\tablenotemark{c} & $10^{-5}(<10^{-5})$ 
& $10^{-4}(<10^{-5})$ & $10^{-3}(<10^{-5})$ & $10^{-3}(<10^{-5})$ \nl
$\left(E_{\nu,\bar{\nu}}/M_{\rm axis}c^2 \right)$\tablenotemark{d} & 
$>5\times10^{-4}$ & $>1.5\times10^{-3}$ & $>2.5$ & $>25$ \nl

\tablenotetext{a}{$E_{\nu,\bar{\nu}}^{\rm max}$ is the neutrino/anti-neutrino 
annihilation energy released by setting the accretion rate 
equal to the mass-transfer rate.  The luminosity is a function of 
both the accretion rate and the spin parameter $a$ (see Eq. 9) and 
we estimate the energy by integrating the luminosity assuming 
the black hole is not spinning before accretion sets in.  
The accretion rate can never exceed the mass-transfer rate, and 
hence, these energies are rough upper limits for the gamma-ray 
burst energy.}
\tablenotetext{b}{We use the magnetic field estimate 
of Popham, Woosley, \& Fryer (1998) which assumes that 
the magnetic field energy is 10\% of the equipartition energy, 
or $0.1 \rho v^2$.}
\tablenotetext{c}{This corresponds to the mass from the 
white dwarf which lies along the accretion disk axis 
and is likely to be swept up in the explosion for a 
beaming factor of 10(100).}
\tablenotetext{d}{$E_{\nu,\bar{\nu}}/M_{\rm axis}c^2 \approx \gamma$ 
when $E_{\nu,\bar{\nu}}/M_{\rm axis}c^2 \gg 1$.  We assume 
beaming factors of 100.}

\enddata
\end{deluxetable}
\newpage

\begin{figure}
\plotfiddle{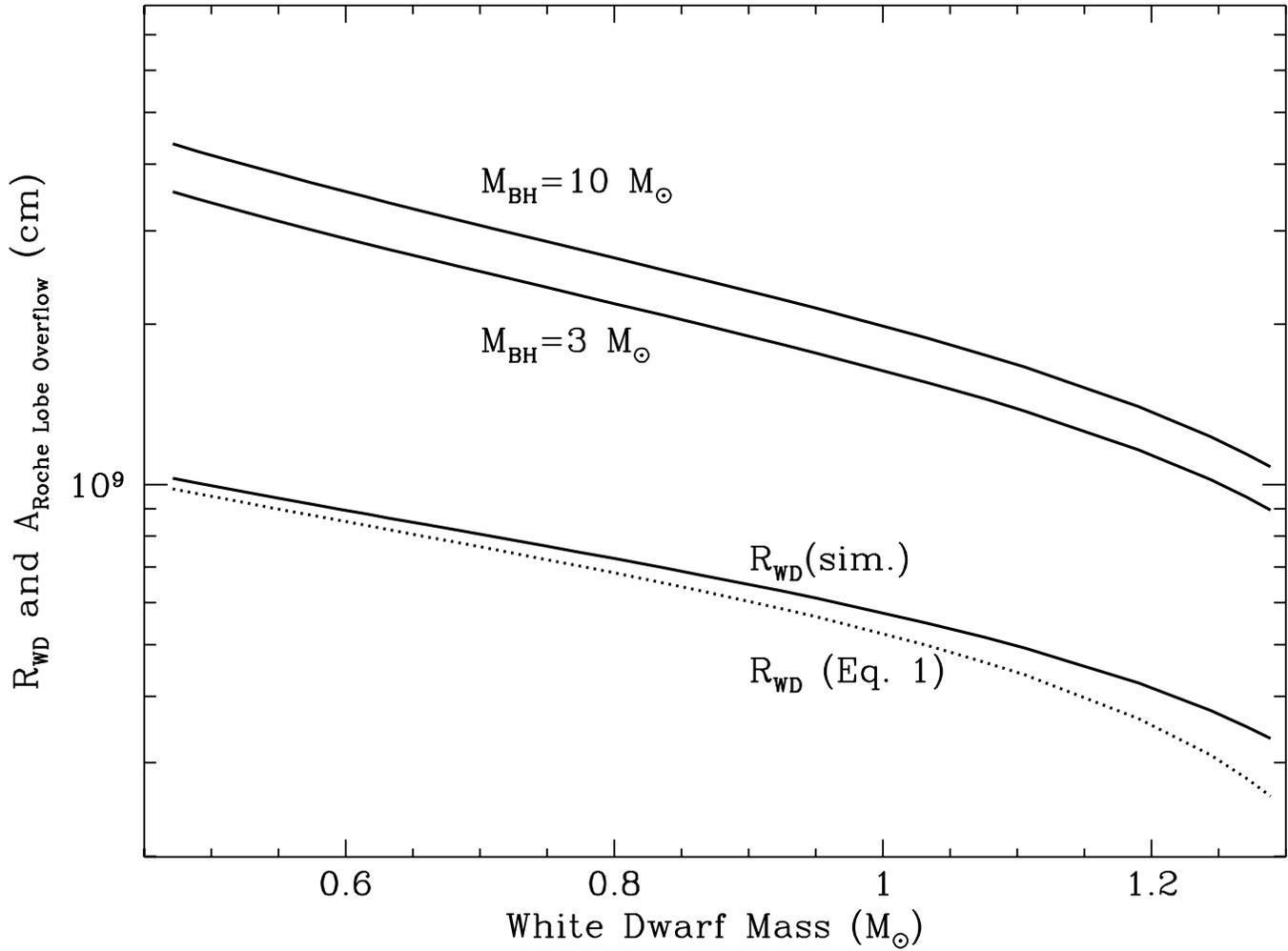}{7in}{-90}{70}{70}{-280}{520}
\caption{White dwarf radii and separations at which 
Roche-lobe overflow occurs vs. white dwarf mass.  The 
dotted line shows the white dwarf radius using eq. (\ref{eq:rwd}) 
in comparison to our simulated radii.  The orbital separations 
are given for two black hole masses ($3,10$\,M\sun) and 
demarkate the limit within which the white dwarf overfills its 
Roche lobe.}
\end{figure}

\begin{figure}
\plotfiddle{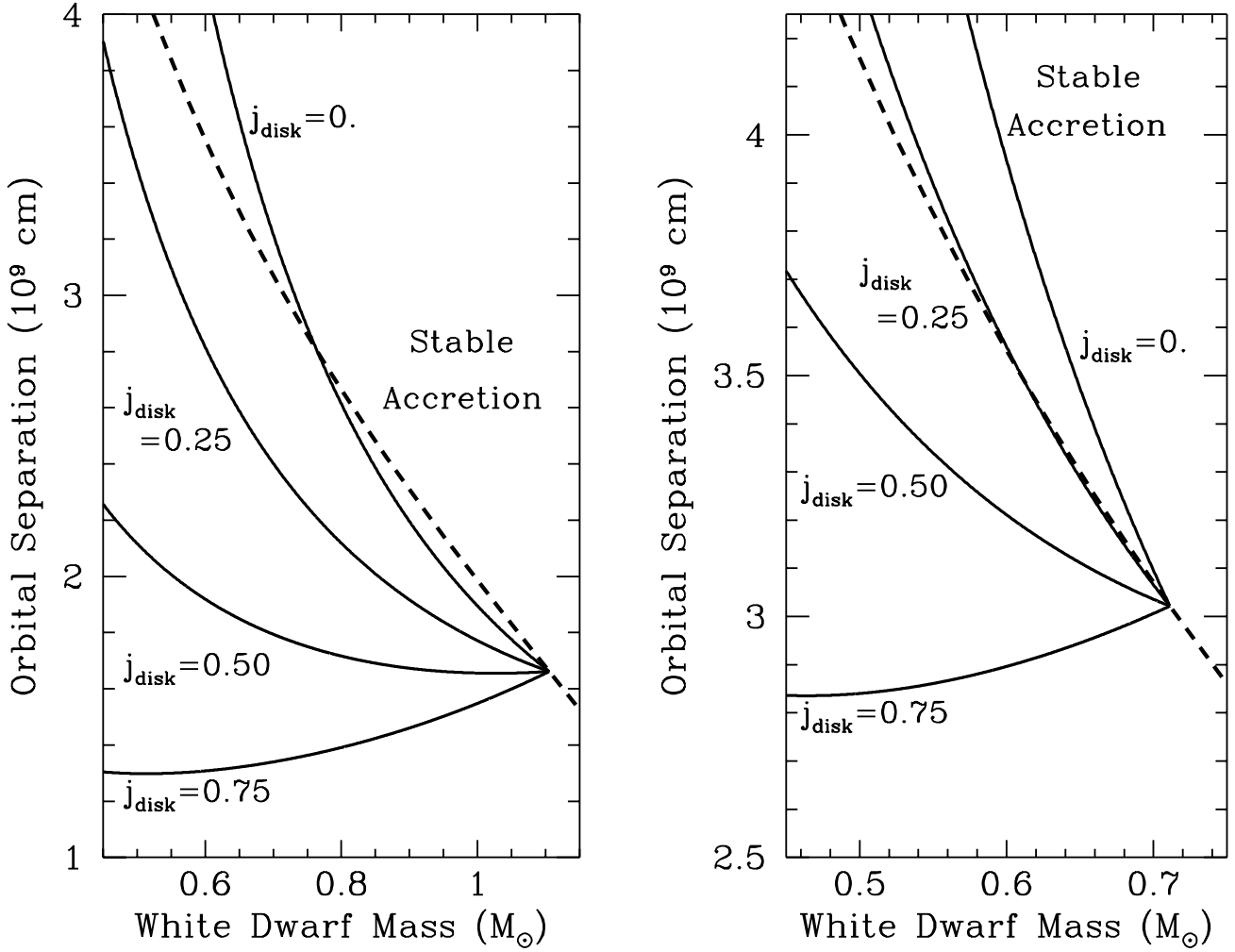}{7in}{-90}{70}{70}{-280}{520}
\caption{Evolutionary paths of the binary separation 
as a white dwarf accretes onto a 3\,M\sun black hole for 
a range of $j_{\rm disk}$ (fraction of specific 
angular momentum of the the white dwarf) values.  The critical 
Roche-lobe separation is plotted for comparison.  
If the separation remains above this critical separation, 
stable accretion occurs.  Otherwise, the accretion 
is unstable and the white dwarf quickly accretes onto 
the black hole.}
\end{figure}

\begin{figure}
\plotfiddle{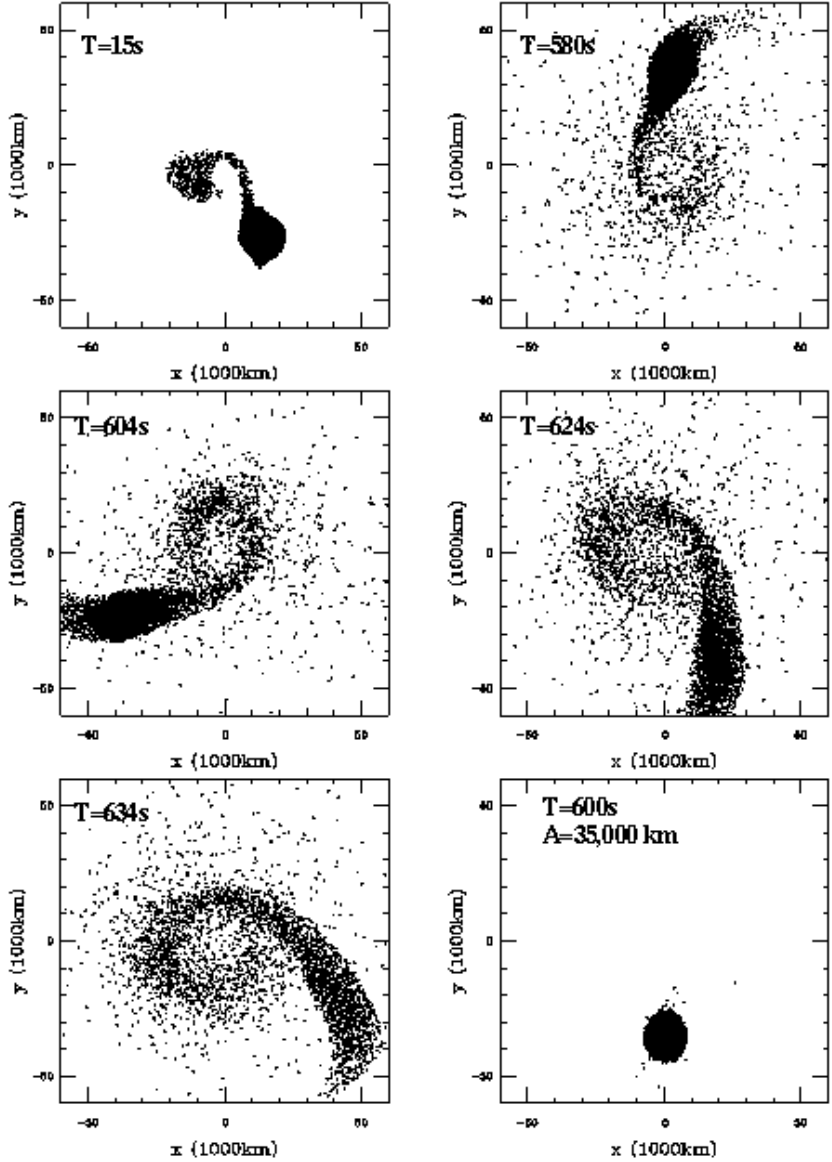}{7in}{0}{70}{70}{-220}{0}
\caption{Time evolution of a $M_{\rm WD}=0.7 {\rm M_\odot},
M_{\rm BH}=3.0 {\rm M_\odot}$ simulation.  Here we show slices 
about the z-axis from -10000-10000\,km.  Note that very 
little accretion occurs for over 600s and then, very 
rapidly, the white dwarf is torn apart.  However, for 
orbital separations just 20\% further out (lower right 
panel), there is no mass transfer.  This suggests 
our initial conditions are roughly accurate.}
\end{figure}

\begin{figure}
\plotfiddle{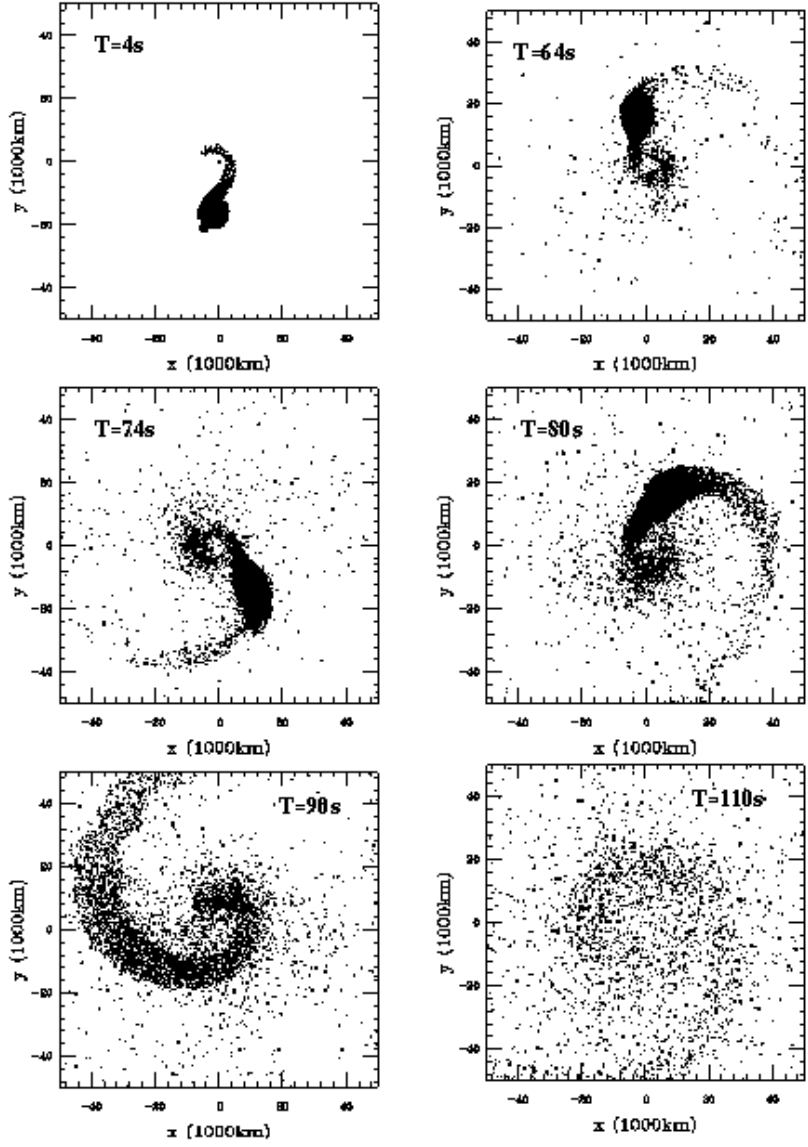}{7in}{0}{70}{70}{-220}{0}
\caption{Time evolution of a $M_{\rm WD}=1.1 {\rm M_\odot},
M_{\rm BH}=3.0 {\rm M_\odot}$ simulation.  Here we show slices 
about the z-axis from -10000-10000\,km.  Note that very 
little accretion occurs for over 70s and then, very 
rapidly, the white dwarf is torn apart.  In the last 
slide (T=110s), nearly all traces of the white dwarf 
have been removed and half of the white dwarf mass 
has been accreted onto the black hole.}
\end{figure}

\begin{figure}
\plotfiddle{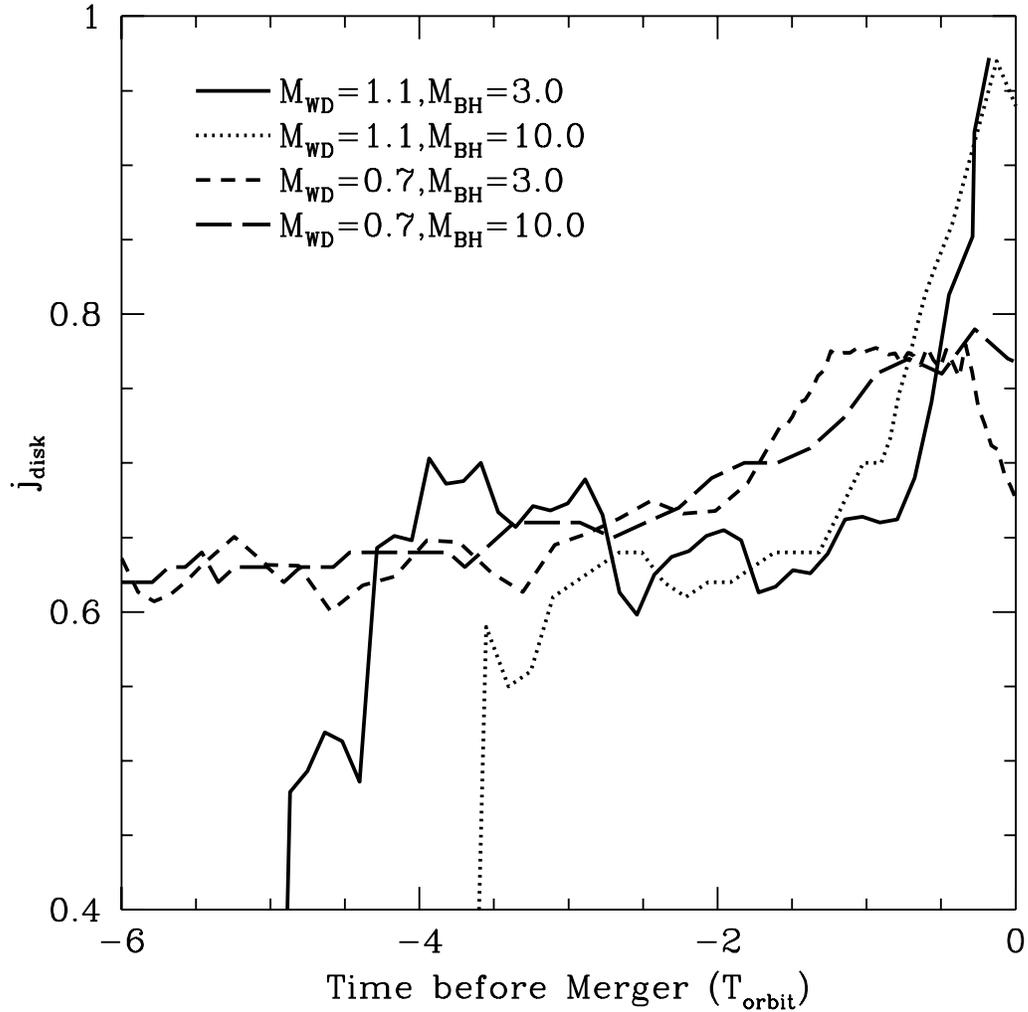}{7in}{0}{70}{70}{-220}{0}
\caption{The specific angular momentum ($j_{\rm disk}$) 
of the disk (in units of the white dwarf specific angular 
momentum) as a function of time (in orbital time).  The 
orbital times of the systems are given in Table 1.  As 
the white dwarf is disrupted, its entire angular momentum 
is put in the disk.}
\end{figure}

\begin{figure}
\plotfiddle{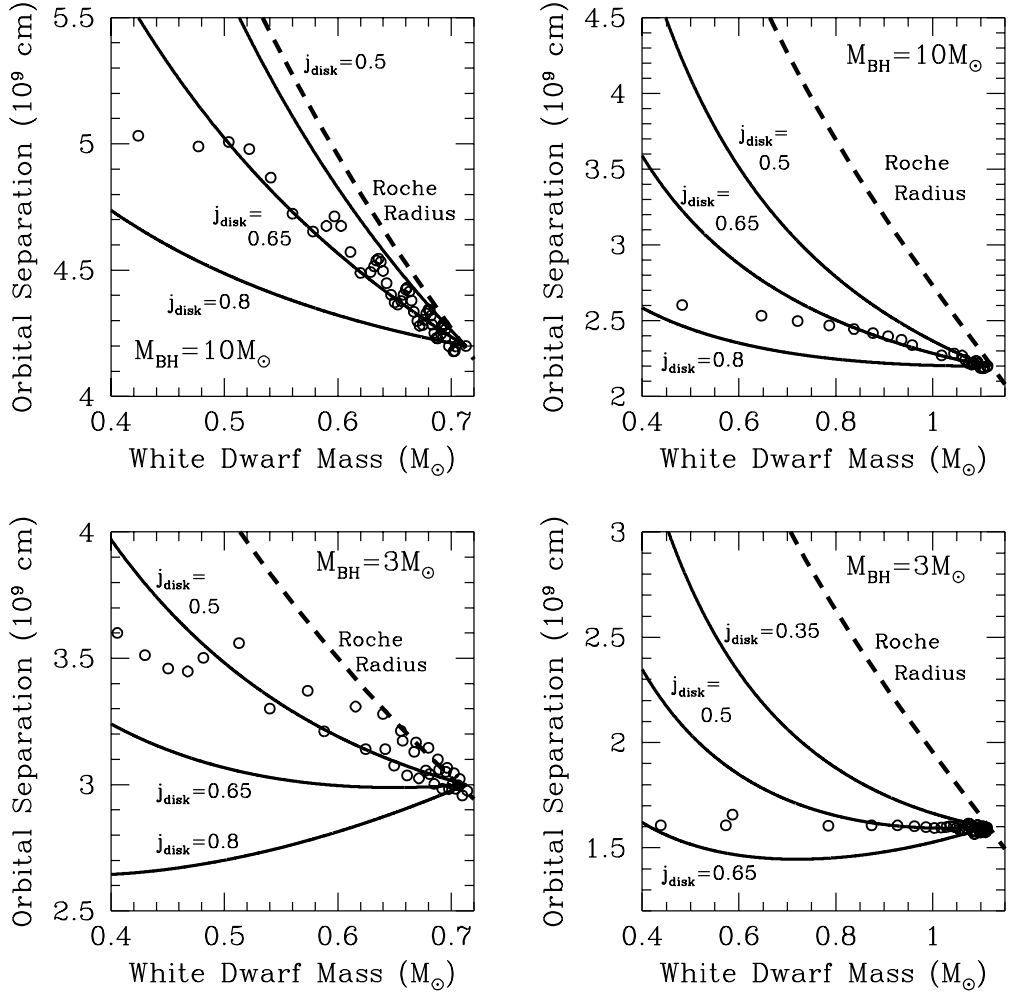}{7in}{0}{70}{70}{-210}{0}
\caption{Orbital separation as a function of white dwarf 
mass.  The circles are the data from the simulations and 
the solid lines denote the predicted separations from 
equation (\ref{eq:rad}).  The dashed line is the separation 
at which the white dwarf overfills its Roche radius.  
Note that after losing $\sim 0.2$M\sun, the orbital 
separation evolution remains constant.  This occurs as 
the white dwarf is torn apart by tidal forces.}
\end{figure}

\begin{figure}
\plotfiddle{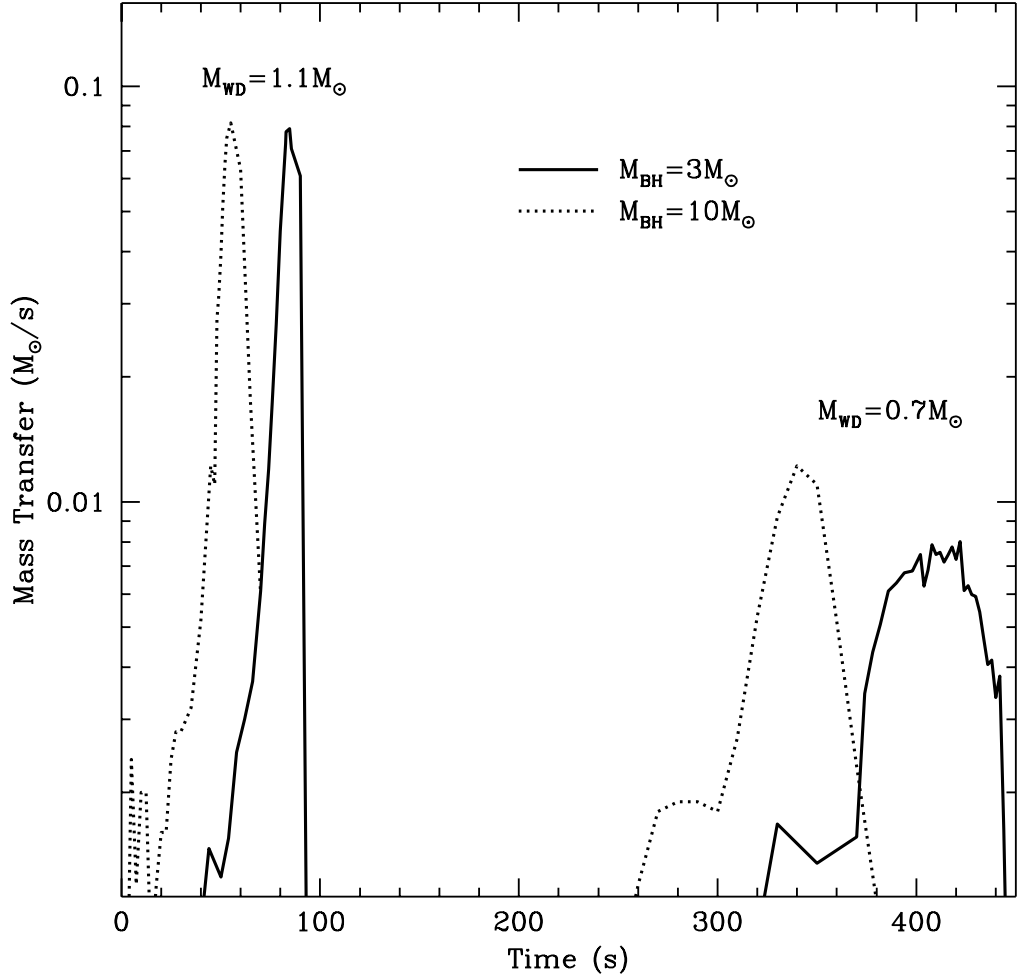}{7in}{0}{70}{70}{-220}{0}
\caption{Mass Transfer Rate vs. time for the 4 
white dwarf/black hole mergers.  The more massive 
white dwarfs merge more quickly and have correspondingly 
higher mass-transfer rates.  This rate gives a maximum 
for the disk accretion rate onto the black hole.}
\end{figure}

\begin{figure}
\plotfiddle{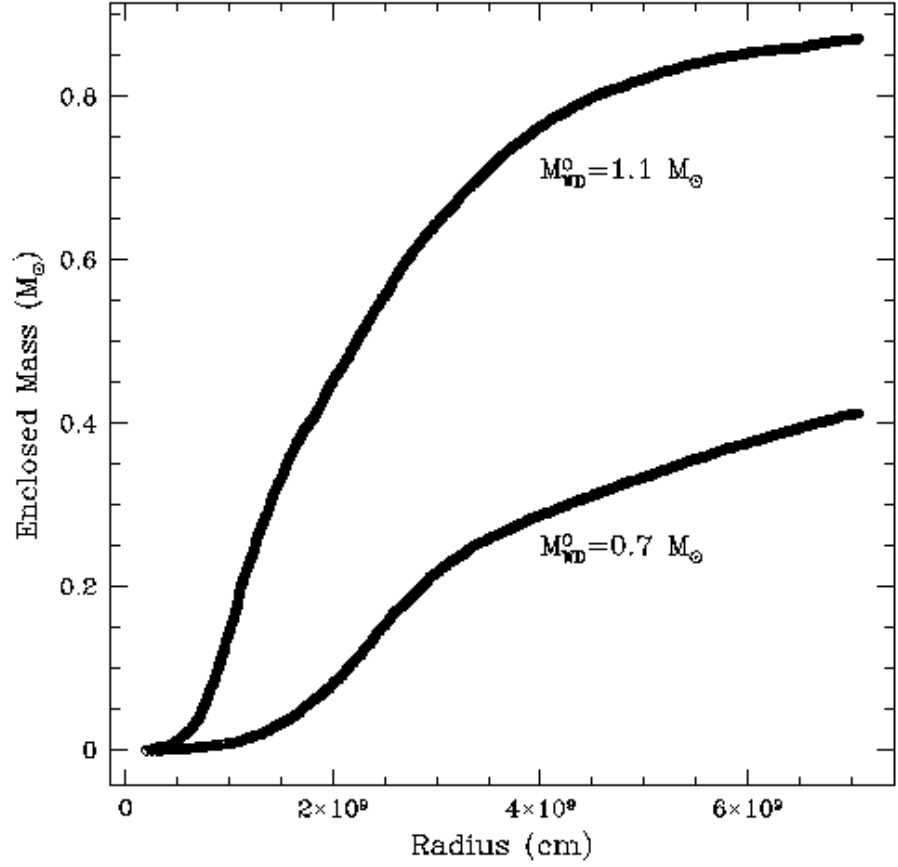}{7in}{0}{70}{70}{-220}{0}
\caption{Enclosed Mass vs. radius just after white 
dwarf disruption.  The accretion rate of this material 
onto the disk is roughly given by:  $\dot{M}_{\rm acc}
\approx \alpha \Omega M_{\rm disk}$.}  
\end{figure}

\begin{figure}
\plotfiddle{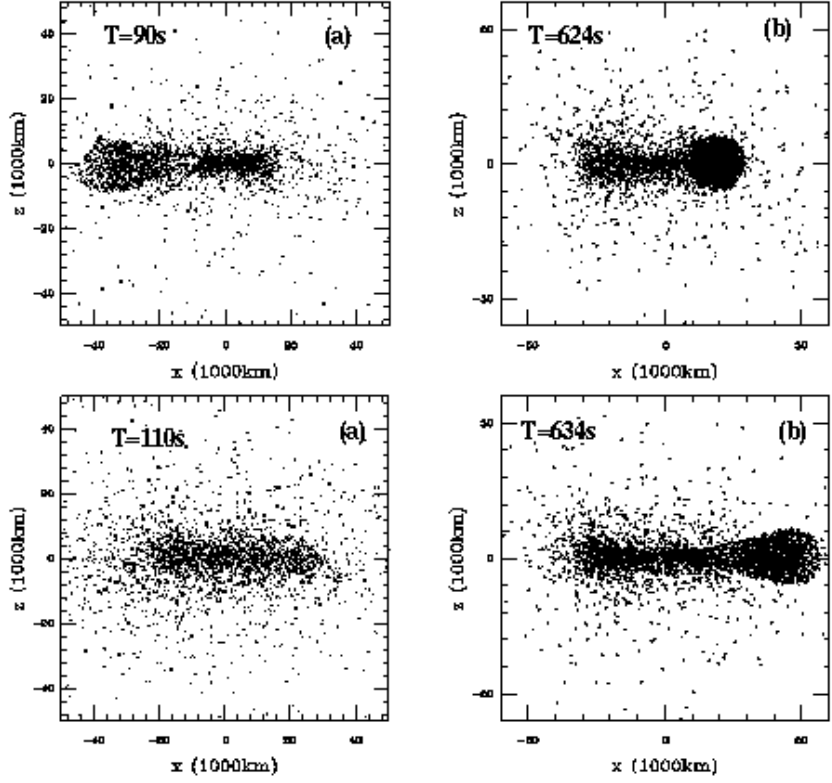}{7in}{0}{70}{70}{-220}{0}
\caption{Distribution of the white dwarf material 
along the rotation axis for our two $M_{\rm BH}=3$M\sun 
simulations.  Any GRB explosion must plow through this 
material (and possibly sweep it up) as it expands.  
If the explosion is beamed, the total swept up mass 
is very small (see Table 2) and will not effect the 
GRB.}  
\end{figure}

\end{document}